\documentclass[aps,prd,twocolumn,nofootinbib]{revtex4-1}
\usepackage{epsfig}
\usepackage[colorlinks,linkcolor=blue,anchorcolor=blue,citecolor=blue,urlcolor=blue,breaklinks=true]{hyperref}
\usepackage{graphics}
\usepackage{slashed}
\usepackage{color}
\usepackage{amsmath}
\usepackage{bm}
\usepackage{setspace}
\usepackage{multirow}
\usepackage{graphicx}

\begin{document}
\author{Cheng-Ming Li$^{1}$}\email{licm@zzu.edu.cn}
\author{Shu-Yu Zuo$^{2}$}
\author{Ya-Peng Zhao$^{3}$}\email{zhaoyapeng2013@hotmail.com}
\author{Hui-Jun Mu$^{1}$}
\author{Yong-Feng Huang$^{4}$}\email{hyf@nju.edu.cn}
\address{$^{1}$ School of Physics and Microelectronics, Zhengzhou University, Zhengzhou 450001, China}
\address{$^{2}$ College of Science, Henan University of Technology, Zhengzhou 450000, China}
\address{$^{3}$ School of Mathematics and Physics, Henan Urban Construction University, Pingdingshan 467036, China}
\address{$^{4}$ School of Astronomy and Space Science, Nanjing University, Nanjing 210023, China}
\title{The study of nonstrange quark stars within a modified NJL model}

\begin{abstract}
In this work, a modified Nambu-Jona-Lasinio (NJL) model with
proper-time regularization is employed to study the structure of
nonstrange quark stars. The coupling constant of four-fermion
interaction in the conventional NJL model is modified as
$G=G_1+G_2\langle\bar{\psi}\psi\rangle$ to highlight the feedback
of quark propagator to gluon propagator. To study the dependence
of the equation of state (EOS) on this modification as well as the
vacuum pressure, we choose nine representative EOSs for
comparison. It is found that a smaller $G_1$ leads to a stiffer
EOS, and a higher vacuum pressure (i.e., a smaller bag constant)
yields a softer EOS at low energy density. It is further shown
that the heaviest quark star under this modified NJL model
satisfies not only the recent mass measurement of PSR J0740+6620,
but also the radius constraints from X-ray timing observations.
The corresponding tidal deformability is also in agreement with
the observations of GW170817.

\bigskip

\bigskip


\end{abstract}

% QED3 11.10.Kk, 11.15.Tk, 11.30.Qc
% quark-gluon, 12.38.Mh; Quark models, 12.39.-x; Quark confinement, 12.38.Aw; Quark deconfinement, 25.75.Nq; Lattice QCD calculations, 12.38.Gc;
% Quark-gluon plasma, 12.38.Mh; phase transitions in QGP, 25.75.Nq; production of QGP, 25.75.Nq; particle physics of chirality, 11.30.Rd;

\maketitle

\section{INTRODUCTION}

The structure of neutron stars and quark stars is largely
determined by the equation of state (EOS) of dense matter. Given a
particular EOS, the corresponding mass-radius ($M$-$R$) and
mass-central energy density ($M$-$\epsilon_{\rm c}$) relations can
be obtained by solving the Tolman-Oppenheimer-Volkoff (TOV)
equation. Since neutron stars and quark stars are composed of
strongly interacting matter at high densities under a relatively
low temperature, nonperturbative quantum chromodynamics (QCD) is
deeply involved in exploring the EOS and structure of these
compact stars. There are two critical feactures in QCD, i.e.,
color confinement and dynamical chiral symmetry
breaking. At low chemical potentials, quarks are confined in
hadrons under a low temperature. However, at high chemical
potentials, quarks become deconfined. As a result, the observed
highly compact pulsars could be quark stars rather than normal
neutron stars.

In light of the hypothesis that strange quark
matter might be the ground state of strongly interacting
matter~\cite{10.1143/PTP.44.291,terazawa1979tokyo,PhysRevD.4.1601,Witten1984rs},
many authors have extensively studied the characteristics of
strange quark stars, either pure quark stars or hybrid neutron
stars with a quark
core~\cite{PhysRevD.30.2379,PhysRevD.95.056018,PhysRevD.97.103013,
PhysRevD.99.043001,PhysRevD.101.063023,li2021strange,
SEDAGHAT2022137032,GENG2021100152}.
Interestingly, a recent study~\cite{PhysRevLett.120.222001} shows
that stable quark matter might not be strange so that nonstrange
quark stars can exist. Further studies have been
carried out based on this
viewpoint~\cite{PhysRevD.100.123003,PhysRevD.100.043018,
PhysRevD.101.043003,PhysRevD.102.083003,PhysRevD.103.063018}.
In Refs.~\cite{PhysRevD.100.123003,PhysRevD.100.043018}, a new
self-consistent mean-field approximation is employed to study the
properties of nonstrange quark stars, such as the $M$-$R$ relation
and the tidal deformability. Note that the
difference between Ref.~\cite{PhysRevD.100.123003} and
Ref.~\cite{PhysRevD.100.043018} is that the authors of
Ref.~\cite{PhysRevD.100.123003} adopted the proper-time
regularization, while a three-momentum cutoff regularization is
used in Ref.~\cite{PhysRevD.100.043018}. A recent
study~\cite{PhysRevD.105.123004} demonstrated that both the
nonstrange and strange quark matter could be absolutely stable
under the combination of the quark vector interaction and exchange
interaction. Therefore, for two-flavor quark matter and
three-flavor quark matter, which one is more stable is still an
open question at present. In this study, we will investigate the
properties of nonstrange quark stars in depth, providing useful
constraints the EOS of nonstrange quark matter with recent
astronomical observations.

Since Joycelyn Bell and Antony Hewish discovered PSR B1919+21
in 1967, a large number of pulsar mass measurements have been
obtained till now. Among these measurements, PSR J0348+0432
and PSR J0740+6620 are two special examples characterized by
their large masses, i.e., $2.01\pm0.04$ $M_{\odot}$ (solar
mass)~\cite{Antoniadis1233232} and $2.14^{+0.10}_{-0.09}$
$M_{\odot}$~\cite{cromartie2020relativistic}, respectively.
In recent years, the radii of a few pulsars were also
encouragingly measured at unprecedented precision due to
successful operation of the Neutron Star Interior Composition
Explorer (NICER)~\cite{bogdanov2019constraining,capano2020stringent,riley2019nicer}.
For example, in Ref.~\cite{bogdanov2019constraining}, the
radius of a typical 1.44 $M_{\odot}$ pulsar is found to be
larger than 10.7 km. Additionally, the recently discovered
gravitational wave (GW) event GW170817 has opened a new era
of multi-messenger astronomy~\cite{PhysRevLett.119.161101,
2041-8205-848-2-L12,PhysRevLett.120.172703,PhysRevLett.120.172702,
2041-8205-850-2-L19,2041-8205-850-2-L34,PhysRevD.97.084038,
PhysRevD.96.123012,PhysRevD.97.083015,PhysRevD.97.021501,
2041-8205-852-2-L29,2041-8205-852-2-L25,0004-637X-857-1-12,
2018ApJ...862...98Z,2018ApJ...860...57A,ma2019pseudoconformal},
and the LIGO-VIRGO collaboration provided useful constraints
on the dimensionless tidal deformability ($\Lambda$) of neutron
stars through GW waveform observations during the inspiral phase
of the binary neutron star (BNS) merger. For the star of
low-spin priors, it is estimated
as $\Lambda(1.4 M_{\odot})\leq800$~\cite{PhysRevLett.119.161101},
and the $\Lambda_1-\Lambda_2$ relation of GW170817 is also
constrained by considering particular waveform models, such
as the TaylorF2 waveform~\cite{PhysRevX.9.011001}.

Lattice regularized QCD calculations are troubled
by the famous ``sign problem'' at finite chemical potentials,
making it difficult to perform calculations based on the first
principles. As a result, we have to resort to some effective
models to calculate the EOS of quark matter. Astronomical
observations are then used to test these hypothetical EOSs. In
general, the EOS should not be too soft since it will fail to
produce the massive stars of $\sim 2 M_{\odot}$ as observed. At
the same time, the EOS also should not be too stiff when the upper
limit of the tidal deformability is considered, as hinted by the
observational results of GW170817. 

In this study we use a modified
Nambu-Jona-Lasinio (NJL) model to study the structure of strange
stars. Inspired by the operator product expansion (OPE) approach,
the traditional constant coupling coefficient of the 2-flavor NJL
model is modified as $G = G_1 + G_2 \langle \bar{\psi}\psi\rangle$
with $G_2 \langle \bar{\psi}\psi\rangle$ accounting for the
feedback of the quark propagator to the gluon propagator (see for
instance Ref.~\cite{PhysRevD.85.034031,Cui2013,
Cui2014,PhysRevD.93.036006,PhysRevD.94.096003,doi:10.1142/S0217732317501073,
Fan_2019,PhysRevD.97.103013}). The EOS of nonstrange quark matter
derived in this framework will be used to study the properties of
quark stars, especially the $M-R$ relation and the tidal
deformability.

Many studies have focused on the NJL model with
't Hooft
interaction~\cite{PhysRevD.95.056018,PhysRevD.105.123004}.
Although our gap equation in the SU(2) case looks similar to
previous SU(3) studies which possesses a quadratic dependence on
the quark condensate, we would like to point out that the reason
is different here. In
Ref.~\cite{PhysRevD.95.056018,PhysRevD.105.123004}, the t' Hooft
interaction term leads to the above quadratic dependence. But in
this study, the modification of the coupling constant $G$ is
responsible for the effect.

This paper is organized as follows. In Section~\ref{one}, a brief
introduction on the modified 2-flavor NJL model is presented, and
nine representative EOSs for nonstrange quark matter are derived.
In Section~\ref{two}, the tidal deformability and the $M - R$,
$M - \epsilon_{\rm c}$ relations are calculated for quark stars,
and compared with astronomical observations. Finally, a brief
summary and discussion is presented in Section~\ref{three}.

\section{EOS of nonstrange quark matter}\label{one}

The NJL model is widely used as an effective model to describe cold
dense quark matter in neutron stars and quark
stars~\cite{RevModPhys.64.649,Buballa2005205}. The general form of
the Lagrangian for the 2-flavor NJL model is:
\begin{equation}\label{genLag}
\mathcal{L}=\bar{\psi}(i{\not\!\partial}-m) \psi + G[(\bar{\psi}\psi)^2 + (\bar{\psi}i\gamma^5\tau\psi)^2],
\end{equation}
where $m$ denotes the current quark mass, and $G$ is the four-fermion
coupling constant\footnote{An exact isospin symmetry between u and d
quark is adopted in this work so that $m_{\rm u}=m_{\rm d}=m$.}. The
interaction term $G[(\bar{\psi}\psi)^2 + (\bar{\psi}i\gamma^5\tau\psi)^2]$
includes the scalar-scalar and pseudoscalar-isovector channels.

In general, the effective quark mass
$m_{\rm{eff}}$ can be obtained by the self-consistent gap equation
of
\begin{equation}\label{gapeq}
m_{\rm{eff}}=m-2G\langle\bar{\psi}\psi\rangle,
\end{equation}
where $\langle\bar{\psi}\psi\rangle$ is the quark condensate. At
a zero temperature and zero chemical potential, it can be calculated as
\begin{eqnarray}
  \langle\bar{\psi}\psi\rangle & = & -\int\frac{{\rm d}^4p}{(2\pi)^4}{\rm Tr}[iS(p^2)]\nonumber\\
  &=& -N_{\rm c}\int_{-\infty}^{+\infty}\frac{{\rm d}^4p}{(2\pi)^4}\frac{8im_{\rm{eff}}}{p^2-m_{\rm{eff}}^2},\,\,\label{qcondensate}
\end{eqnarray}
where the trace ``Tr'' is evaluated in color, flavor, and Dirac spaces,
and $S(p^2) = \frac{1}{\not p-m_{\rm{eff}}}$ represents the quark propagator.

To proceed, we need to convert our equations from the Minkowski
space to the Euclidean space and employ some kinds of regularization.
In this study, we adopt the proper-time regularization (PTR), a covariant regularization that has
a ``soft'' cutoff to avoid the ultraviolet divergence when the
momentum integration is to infinity. The formula of PTR is
\begin{eqnarray}
  \frac{1}{X^n} &=& \frac{1}{(n-1)!}\int_{0}^{\infty}{\rm d}\tau\tau^{n-1}e^{-\tau X}\nonumber \\
   & &\xrightarrow{\rm{UV\,\,cutoff}} \frac{1}{(n-1)!}\int_{\tau_{\rm UV}}^{\infty}{\rm d}\tau\tau^{n-1}e^{-\tau X},\,\,\label{sregularization}
\end{eqnarray}
where the integral limit $\tau_{\rm UV}$ is related to the
ultraviolet cutoff $\Lambda_{\rm UV}$: $\tau_{\rm UV} =
\Lambda^{-2}_{\rm UV}$. Adopting the PTR regularization,
Eq.~(\ref{qcondensate}) becomes
\begin{eqnarray}
   \langle\bar{\psi}\psi\rangle &=& -N_{\rm c}\int_{-\infty}^{+\infty}\frac{{\rm d}^4p^{\rm E}}{(2\pi)^4}\frac{8im_{\rm{eff}}}{(p^{\rm E})^{2}+m_{\rm{eff}}^2}\nonumber\\
   &=& -\frac{N_{\rm c}}{(2\pi)^4}\int_{-\infty}^{+\infty}\int_{-\infty}^{+\infty}{\rm d}^3\overrightarrow{p}{\rm d}p_4\frac{8m_{\rm{eff}}}{p_4^2+\overrightarrow{p}^2+m_{\rm{eff}}^2}\nonumber \\
  &=& -\frac{6m_{\rm{eff}}}{\pi^2}\int_{0}^{+\infty}{\rm d}p\frac{p^2}{\sqrt{p^2+m_{\rm{eff}}^2}}\nonumber \\
   &=& -\frac{6m_{\rm{eff}}}{\pi^{\frac{2}{5}}}\int_{\tau_{\rm UV}}^{\infty}\int_{0}^{+\infty}{\rm d}\tau {\rm d}p\tau ^{-\frac{1}{2}}p^2e^{-\tau (m_{\rm{eff}}^2+p^2)}\nonumber \\
   &=& -\frac{6m_{\rm{eff}}}{4\pi^2}\int_{\tau_{\rm UV}}^{\infty}{\rm d}\tau \frac{e^{-\tau m_{\rm{eff}}^2}}{\tau^2},\,\,\label{regofqcondensate}
\end{eqnarray}
where the superscript $E$ means the parameter is measured
in the Euclidean space.

According to the NJL theory, the coupling constant $G$ represents
the effective gluon propagator. Note that the quark and gluon
propagators satisfy different Dyson-Schwinger (DS) equations, but
they should couple with each other in view of QCD theory. As we
know, quark propagators in the Nambu phase and Wigner phase are
very different from each other
~\cite{Cui2018,0954-3899-45-10-105001,PhysRevD.99.076006}, so the
corresponding gluon propagators in these two phases should also be
different. However, in the normal NJL model, $G$ is simplified as
a constant and remains the same in both the Nambu phase and the
Wigner phase, which is obviously unreasonable. Additionally,
simulations of lattice QCD have shown that the gluon propagator
should vary with temperature, while its dependence on the chemical
potential is more uncertain. In most NJL
calculations, the effective gluon propagator $G$ is usually
assumed to be ``static'', which thus does not depend on the
temperature and chemical potential. It is obviously in
contradiction with the requirements exerted on an effective gluon
propagator.

The plane wave method in the QCD sum rule
approach is used in Ref.~~\cite{REINDERS19851}. It is argued that
the full Green function (which is unknown) can be divided into two
parts: the nonperturbative part and the perturbative part. The
condensates are then expressed as various moments of the
nonperturbative Green function. Therefore, the most general form
of the ``nonperturbative'' gluon propagator is
\begin{equation}\label{gapeq}
D_{\mu\nu}^{\rm{npert}}\equiv D_{\mu\nu}^{\rm{full}}-D_{\mu\nu}^{\rm{pert}}\equiv c_1\langle\bar{\psi}\psi\rangle+c_2\langle G^{\mu\nu}G_{\mu\nu}\rangle+...,
\end{equation}
where $\langle G^{\mu\nu}G_{\mu\nu}\rangle$ is the gluon
condensate, $c_1$ and $c_2$ are coefficients that can be
calculated with the QCD sum rule
approach~\cite{Steele1989,pascual1984qcd}, and the ellipsis refers
to the contributions from other condensates (e.g., the mixed
quark-gluon condensate). 

Among all the condensates, the quark condensate
has the lowest dimension, and a nonzero value of which, in the
chiral limit, just signals the dynamical chiral symmetry breaking.
Therefore, it is the elementary item and plays the most important
role in the QCD sum rule approach. In this study, we will treat
its contribution separately, while the contribution of other
condensates is included in the perturbative part of the gluon
propagator. In the framework of the NJL model, it is equivalent to
a modification of
\begin{equation}\label{effG}
G\rightarrow G_1+G_2\langle\bar{\psi}\psi\rangle,
\end{equation}
which is quite similar to the approach in
Refs.~\cite{PhysRevD.85.034031,Cui2013,Cui2014,PhysRevD.93.036006,PhysRevD.94.096003,doi:10.1142/S0217732317501073,Fan_2019,PhysRevD.97.103013}.
Under this modification, the coupling strength $G$ will depend on
both u and d quark condensates. $G_2$ can be regarded as an
effective coupling strength, reflecting the relative weight of the
influence of the quark propagator and gluon propagator
\footnote{In a conventional two-flavor NJL model,
the thermodynamical potential is obtained based on the mean-field
approximation, see e.g. Eq. (2.46) in Ref.~\cite{Buballa2005205}
(neglecting the contribution of the vector interaction). The
modification here can be regarded as a scheme beyond the
mean-field approach, and it is hard to find a closed and easily
tractable effective potential (a more detailed analysis can be
found in Sec. 2 of Ref.~\cite{Cui2013}).}.
%% We hope that such a
%% simple model can help us to capture the essential physics of
%% the gluon propagator and QCD phase transition.}

In this study, we will take three representative sets of ($G_1$,
$G_2$) and then constrain the corresponding EOSs with astronomical
observations (see Sec.~\ref{two} for more details). The current
quark mass is taken as $m = ( m_{\rm u} + m_{\rm d} ) / 2 = 3.5$
MeV ~\cite{14858815b15b4b008bf0d4556a93e7d9}. Similar to
Ref.~\cite{RevModPhys.64.649}, we fix the parameters
($\Lambda_{\rm UV}$, $G$) to reproduce the experimental data
($f_{\pi}=92$ MeV, $m_{\pi}=135$ MeV). The complete parameter sets
adopted, including $G_1$ and $G_2$, are presented in
Table.~\ref{parameters}.

\begin{table}
\caption{Parameters adopted in this study.}\label{parameters}
\begin{tabular}{p{0.9cm} p{0.9cm} p{0.9cm}p{1.2cm}p{1.2cm}p{1.2cm}p{1.2cm}}
%\begin{tabular}{cccccc}
\hline\hline
$\quad m$&$\,\,\Lambda_{\rm UV}$&$\,\,\,\, m_{\rm{eff}}$&$-\langle\bar{\psi}\psi\rangle^{\frac{1}{3}}$&$\quad\,\, G$&$\quad\,\,\, G_1$&$\quad\,\,\, G_2$\\
$[{\rm MeV}]$&$[{\rm MeV}]$&$[{\rm MeV}]$&$\,\,[{\rm MeV}]$&$[{\rm GeV}^{-2}]$&$[{\rm GeV}^{-2}]$&$[{\rm GeV}^{-5}]$\\
\hline
$\,$&$\,$&$\,$&$\,$&$\,$&$\quad1.935$&$\,\,\,-1.582$\\
$\,\,\,\,$3.5&$\,$1324&$\,\,\,180$&$\,\,\,\,\,353$&$\,\,\,2.005$&$\quad2.005$&$\quad\quad$0\\
$\,$&$\,$&$\,$&$\,$&$\,$&$\quad2.100$&$\quad2.161$\\
\hline\hline
\end{tabular}
\end{table}

\begin{figure}
\includegraphics[width=0.47\textwidth]{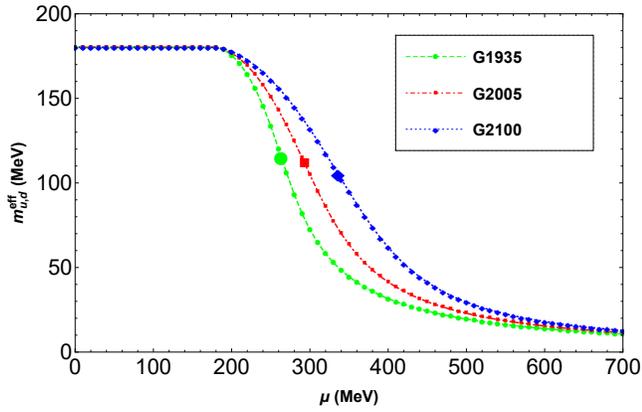}
\caption{The effective quark mass versus $\mu$ at $T=0$. The lines
marked with G1935, G2005, and G2100 correspond to the three
parameter sets in Table.~\ref{parameters}, with $G_1=1.935, 2.005,
2.100$ GeV$^{-2}$, respectively. The
pseudo-critical point is also marked on each line, at $\mu=263,
293, 336$ MeV, respectively.} \label{Fig:Mud}
\end{figure}

\begin{figure}
\includegraphics[width=0.47\textwidth]{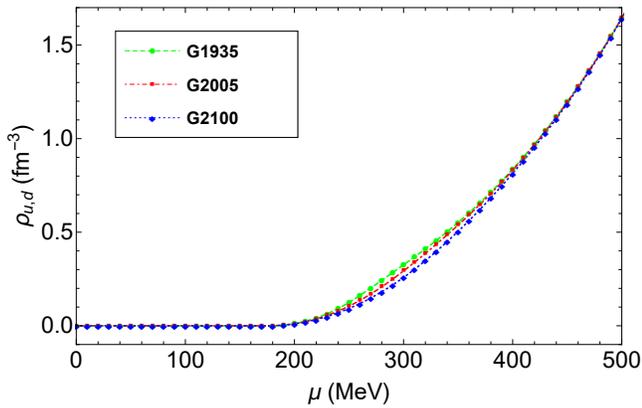}
\caption{The number density of $u, d$ quarks versus $\mu$ at $T=0$
for the three parameter sets in Table.~\ref{parameters}. Line
styles are the same as those in Fig.~\ref{Fig:Mud}.}
\label{Fig:nud}
\end{figure}

Here, we will also extend our calculations to zero temperature and
finite chemical potential, which is equivalent to perform a
transformation in the Euclidean space as~\cite{PhysRevC.71.015205}
\begin{equation}\label{muinpfour}
  p_4\rightarrow p_4+i\mu .
\end{equation}
The quark condensate and number density can then be derived as
\begin{eqnarray}
  \langle\bar{\psi}\psi\rangle&=&-N_{\rm c}\int_{-\infty}^{+\infty}\frac{{\rm d}^4p}{(2\pi)^4}\frac{8m_{\rm{eff}}}{(p_4+i\mu)^2+m_{\rm{eff}}^2+\overrightarrow{p}^2}\nonumber \\
   &=& -\frac{6m_{\rm{eff}}}{\pi^3}\int_{0}^{+\infty}\!{\rm d}p\int_{-\infty}^{+\infty}{\rm d}p_4\frac{p^2}{(p_4+i\mu)^2+m_{\rm{eff}}^2+p^2}\nonumber\\
   &=& \!\left\{
  \begin{array}{lcl}
\displaystyle{\!\!\!-\frac{6m_{\rm{eff}}}{\pi^2}\!\int_{\sqrt{\mu^2-m_{\rm{eff}}^2}}^{+\infty}{\rm d}p\textstyle{\frac{\left[1-{\rm Erf}(\sqrt{m_{\rm{eff}}^2+p^2}\sqrt{\tau_{\rm UV}})\right]p^2}{\sqrt{m_{\rm{eff}}^2+p^2}}}}, m_{\rm{eff}}<\mu\nonumber\\
\displaystyle{\frac{3m_{\rm{eff}}}{2\pi^2}\left[\textstyle{-m_{\rm{eff}}^2{\rm Ei}(-m_{\rm{eff}}^2\tau_{\rm UV})-\frac{e^{-m_{\rm{eff}}^2\tau_{\rm UV}}}{\tau_{\rm UV}}}\right]},\quad\,\,\,\,\!m_{\rm{eff}}>\mu\nonumber
  \end{array}\right.\qquad\qquad
\end{eqnarray}
\begin{equation}\label{qcforTmu}
\quad\quad\quad
\end{equation}
\begin{eqnarray}
  \rho_{\rm i}(\mu) &=& \langle\psi^+\psi\rangle_{\rm i}\nonumber \\
   &=& -N_{\rm c} \int\frac{{\rm d}^4p}{(2\pi)^4}tr\left[iS_{\rm i}\gamma_0\right]\nonumber\\
   &=& 2N_{\rm c}\int\frac{{\rm d}^3p}{(2\pi)^3}\theta(\mu-\sqrt{p^2+m_{\rm{eff}}^2})\nonumber\\
   &=& \left\{
\begin{array}{lcl}
 \frac{1}{\pi^2}(\sqrt{\mu^2-m_{\rm{eff}}^2})^3,             & &\mu>m_{\rm{eff}}\\
  0,                                                & &\mu<m_{\rm{eff}}
   \end{array}
   \right.\label{qnd}
\end{eqnarray}
where the subscript ``i'' denotes the quark of flavor i, and the
trace ``tr'' is calculated in the Dirac space. The effective quark
mass $m_{\rm u,d}^{\rm{eff}}$ as a function of the chemical potential at zero
temperature is shown in Fig.~\ref{Fig:Mud}. The number density
$\rho_{\rm u,d}$ is correspondingly illustrated in
Fig.~\ref{Fig:nud}.

From Fig.~\ref{Fig:Mud}, we can see that for all the three
parameter sets in Table.~\ref{parameters}, there is only crossover
but no any chiral phase transitions as the chemical potential
increases from 0 to 700 MeV. Note that the chemical potential of
the pseudo-critical point is different for the three lines,
as marked in Fig.~\ref{Fig:Mud}. It can also be
seen that at zero temperature and zero chemical potential, the
effective quark masses are the same for the three parameter sets.
This is easy to understand. Under such a condition, $G_1+G_2
\langle\bar{\psi}\psi\rangle$ simply reduces to $G$, which is
exactly the coupling constant of conventional NJL model.
In Fig.~\ref{Fig:nud}, we see that the quark densities differ only in the crossover region of $\mu
\sim (200,400)$ MeV for the three parameter sets.

\begin{figure}
\includegraphics[width=0.47\textwidth]{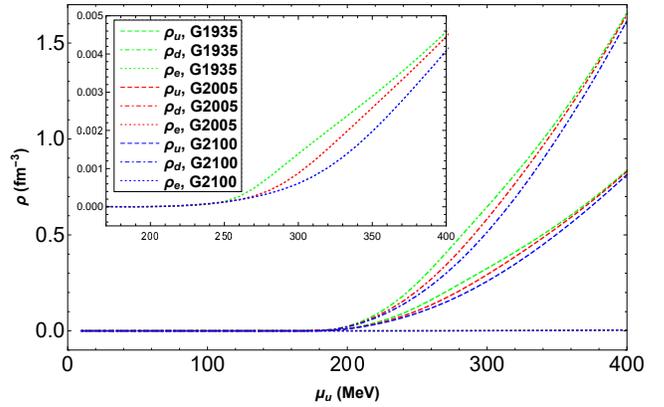}
\caption{The number densities of $u, d$ quarks and electrons
versus $\mu_{\rm u}$ at $T=0$. The matter here is in
beta equilibrium and electric charge neutrality.
The parameter sets are described in Table.~\ref{parameters}. The
inset shows a zoom-in of the electron densities.} \label{Fig:rhoforeq}
\end{figure}

Considering the electrical neutrality of neutron stars and quark
stars as well as electroweak reactions in them, we should take the
beta equilibrium and electric charge neutrality
conditions into account,
\begin{eqnarray}\label{constrains}
           &&\mu_{\rm d}=\mu_{\rm u}+\mu_{\rm e}, \nonumber \\
           &&\frac{2}{3}\rho_{\rm u}-\frac{1}{3}\rho_{\rm d}-\rho_{\rm e}=0,
\end{eqnarray}
where the number density of electrons at zero temperature is
$\rho_{\rm e}(\mu_{\rm e})=\frac{\mu_{\rm e}^3}{3\pi^2}$.
The number densities of quarks and electrons
incorporating these conditions are displayed in
Fig.~\ref{Fig:rhoforeq}. We can see that in each case, the density
of d quarks is larger than that of u quarks when $\mu_u>180$ MeV.
In fact $\rho_d$ approximately equals $2\rho_u$, because the
electron density is much smaller.
Fig.~\ref{Fig:rhoforeq} also shows that as $G_1$
increases, the constituent particle number density decreases for
the same $\mu_{\rm u}$ when $\mu_{\rm u}\geq180$ MeV.
It is well known that under the beta equilibrium
and electric charge neutrality conditions, the quark condensates,
dynamical masses, and densities are different for u quarks and d
quarks. But note that these quantities are connected with each
other, and the relations between them should be considered when
calculating the EOS.

\begin{figure}
\includegraphics[width=0.47\textwidth]{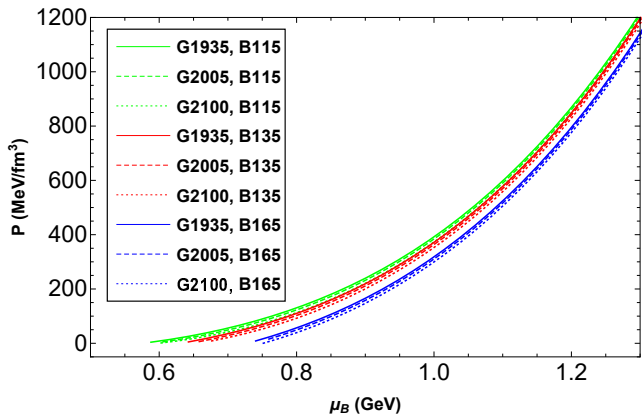}
\caption{The pressure as a function of the baryonic chemical
potential ($\mu_{\rm B}$) for nine representative EOSs at $T=0$.
B115, B135, B165 refers to $B^{\frac{1}{4}}=115, 135, 165$ MeV,
respectively. } \label{Fig:EOSs}
\end{figure}

By definition, the EOS of QCD for $T=0$ and $\mu\neq0$ is~\cite{doi:10.1142/S0217751X08040457}
\begin{equation}\label{EOSofQCD}
  P(\mu)=P(\mu=0)+\int_{0}^{\mu}d\mu'\rho(\mu'),
\end{equation}
where $P(\mu=0)$ represents the negative vacuum pressure and is
independent on the chemical potential. In general, it is treated
as a phenomenological parameter which cannot be calculated
model-independently. The parameter of $P(\mu=0)$ in QCD is
equivalent to the vacuum bag constant ($-B$) of the MIT bag model.
Generally, $B^{\frac{1}{4}}$ should be in a range of $100 - 200$
MeV~\cite{LU1998443,PhysRevD.46.3211}. More specifically, in
Ref.~\cite{PhysRevD.97.083015} and Ref.~\cite{PhysRevD.98.083013},
it is suggested to be $134.1 - 141.4$ MeV and $166.16 - 171.06 $
MeV, respectively. Therefore, in this study, we will choose three
representative values of $B^{\frac{1}{4}}$ to calculate the
nonstrange quark EOS, i.e. $B^{\frac{1}{4}}=115, 135, 165$ MeV.
The results of nine representative EOSs are illustrated in
Fig.~\ref{Fig:EOSs}. We see that the critical baryonic chemical
potential $\mu_{\rm BC}$\footnote{For 2-flavor quark matter, the
baryonic chemical potential is $\mu_{\rm B}=\mu_{\rm u}+\mu_{\rm
d}$.  Note that in Fig.~\ref{Fig:EOSs}, the ``critical point''
refers to the condition that the pressure begins to be nonzero,
not indicating any normal QCD phase transitions.} increases as
$B^{\frac{1}{4}}$ or $G_1$ increases. Additionally, we see that
the EOS is largely determined by the bag constant, while the
parameter of $G_1$ does not affect the EOS significantly.

\begin{figure}
\includegraphics[width=0.47\textwidth]{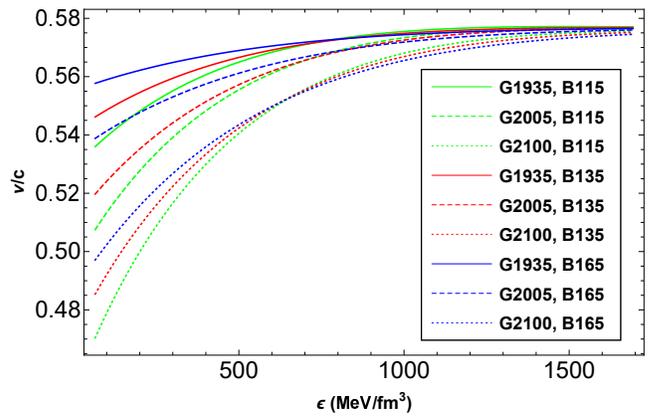}
\caption{The sound velocity versus energy density for the nine
representative EOSs. Line styles are the same as in
Fig.~\ref{Fig:EOSs}. } \label{Fig:sv}
\end{figure}

The relation between the energy density and pressure
is~\cite{PhysRevD.86.114028,PhysRevD.51.1989}
\begin{equation}\label{rbedasp}
  \epsilon=-P+\sum_{i}\mu_{\rm i}\rho_{\rm i}.
\end{equation}
To illustrate the rationality of the nine quark EOSs as well as
their stiffness, we calculate the sound velocity, which is
\begin{equation}\label{soundvelocity}
 \nu = \sqrt{\frac{{\rm d}P}{{\rm d}\epsilon}}.
\end{equation}
The results are shown in Fig.~\ref{Fig:sv}. We see that the sound
velocity does not exceed the conformal limit in all the cases,
i.e., $(\nu/c)^2\leq1/3$, where $c$ is the speed of light.
Usually, a larger $G_1$ leads to a stiffer EOS. Also, note that at
low energy densities, a higher vacuum pressure (i.e., a smaller
$B^{\frac{1}{4}}$) yields a softer EOS, but at high energy
densities, the case is opposite.

\section{Structure of nonstrange quark stars}\label{two}

\begin{figure}
\includegraphics[width=0.47\textwidth]{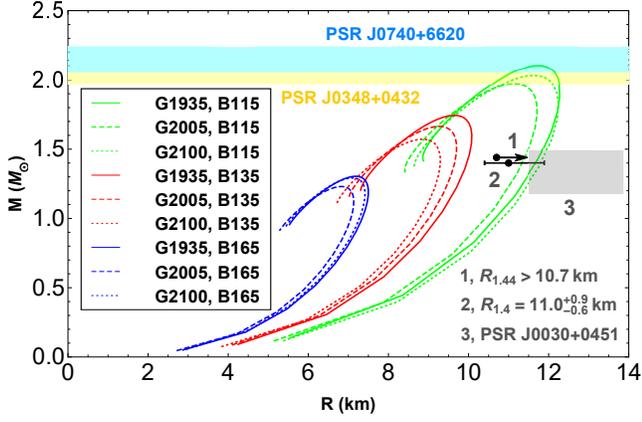}
\caption{$M$-$R$ relations for the nine representative EOSs. The
mass constrains of PSR J0348+0432~\cite{Antoniadis1233232} and PSR
J0740+6620~\cite{cromartie2020relativistic}, and the radius
constraints from NICER x-ray timing
observations~\cite{bogdanov2019constraining,capano2020stringent,riley2019nicer}
are also plot.} \label{Fig:MR}
\end{figure}

\begin{figure}
\includegraphics[width=0.47\textwidth]{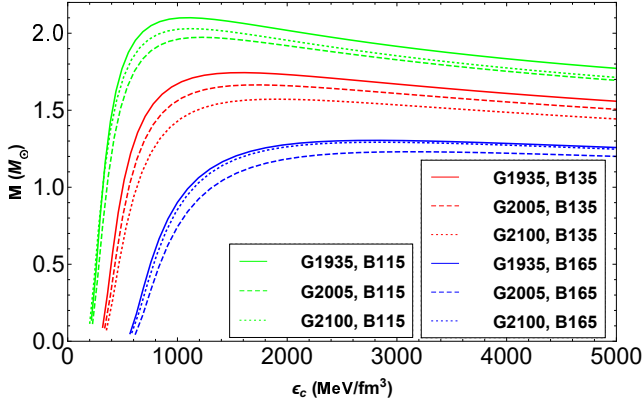}
\caption{$M$-$\epsilon_{\rm c}$ relations for the nine
representative EOSs.} \label{Fig:Mepsilon}
\end{figure}

To study the structure of neutron stars and quark stars, we need
to solve the TOV equation below,
\begin{eqnarray}
  \frac{{\rm d}P(r)}{{\rm d}r} &=& -\frac{G(\epsilon+P)(M+4\pi r^3P)}{r(r-2GM)} \,\, ,\nonumber\\
  &&\frac{{\rm d}M(r)}{{\rm d}r} = 4\pi r^2\epsilon\,\,\, .\label{TOV}
\end{eqnarray}
Using the nine representative EOSs, we have solved the equation
numerically. Our results of $M$-$R$ and $M$-$\epsilon_{\rm c}$
relations are presented in Fig.~\ref{Fig:MR} and
Fig.~\ref{Fig:Mepsilon}, respectively. To constrain the EOSs, we
have also plot some astronomical measurements in
Fig.~\ref{Fig:MR}, including the largest pulsar masses (i.e.,
$2.01\pm0.04$ $M_{\odot}$ for PSR
J0348+0432~\cite{Antoniadis1233232}, and $2.14^{+0.10}_{-0.09}$
$M_{\odot}$ for PSR J0740+6620~\cite{cromartie2020relativistic}),
and the radii measured through NICER x-ray timing observations (1,
$R_{1.44 M_{\odot}}>10.7$ km~\cite{bogdanov2019constraining}; 2,
$R_{1.4 M_{\odot}}=11.0^{+0.9}_{-0.6}$
km~\cite{capano2020stringent}; 3, $M=1.34^{+0.15}_{-0.16}$
$M_{\odot}$ and $R=12.71^{+1.14}_{-1.19}$ km for PSR
J0030+0451~\cite{riley2019nicer}). From Fig.~\ref{Fig:MR}, we see
that only the EOS with $G_1=1.935$ GeV$^{-2}$ and
$B^{\frac{1}{4}}=115$ MeV satisfies all the above constraints, and
the corresponding maximum mass of quark stars is 2.10 $M_{\odot}$,
with a radius of 11.69 km. When
$G_1$ is fixed, the EOS with a smaller bag constant produces a
larger maximum star mass. In fig.~\ref{Fig:Mepsilon}, a higher maximum mass generally corresponds to a smaller central energy density.

We have also calculated the tidal deformability of quark stars. To
do so, we need to solve a set of differential
equations~\cite{PhysRevD.81.123016},
\begin{eqnarray}
  \frac{dH}{dr} &=& \beta,\nonumber\\
  \frac{d\beta}{dr} &=& 2(1-2\frac{m_r}{r})^{-1}H\{-2\pi[5\epsilon+9P+f(\epsilon+P)]\nonumber\\
  &+&\frac{3}{r^2}+2(1-2\frac{m_r}{r})^{-1}(\frac{m_r}{r^2}+4\pi rP)^2\}\nonumber\\
  &+&\frac{2\beta}{r}(1-2\frac{m_r}{r})^{-1}\{\frac{m_r}{r}+2\pi r^2(\epsilon-P)-1\},\,\label{HbetaEq}
\end{eqnarray}
where $P$ and $H(r)$ represent the pressure and the metric
function, respectively, and $f={\rm d}\epsilon/{\rm d}P$.  Let us
further define a parameter ($y$) as $y=R\beta(R)/H(R)-4\pi
R^3\epsilon_0/M$, where $\epsilon_0$ is the energy density at the
surface of the star. Then the dimensionless tidal Love number
$k_2$ for $l=2$ can be calculated as
\begin{eqnarray}
  &k_2&=\frac{8C^5}{5}(1-2C)^2[2+2C(y-1)-y]\nonumber\\
  &\times&\{2C[6-3y+3C(5y-8)]\nonumber\\
  &+&4C^3[13-11y+C(3y-2)+2C^2(1+y)]\nonumber\\
  &+&3(1-2C)^2[2+2C(y-1)-y]ln(1-2C)\}^{-1},\label{tln}
\end{eqnarray}
where $C=M/R$ is the compactness of the star. According to
Ref.~\cite{PhysRevD.81.123016}, the tidal deformability is related
to $k_2$ as
\begin{equation}\label{TD}
  \Lambda=\frac{2}{3}k_2R^5.
\end{equation}

Note that the quark matter in quark stars is in a deconfined
state. It produces a non-negative pressure at the surface. Due to
the negative vacuum pressure in Eq.~(\ref{EOSofQCD}) and the
negative term $-4\pi R^3\epsilon_0/M$ in the formula of $y$, the
quark density should be nonzero at the surface.

\begin{figure}
\includegraphics[width=0.47\textwidth]{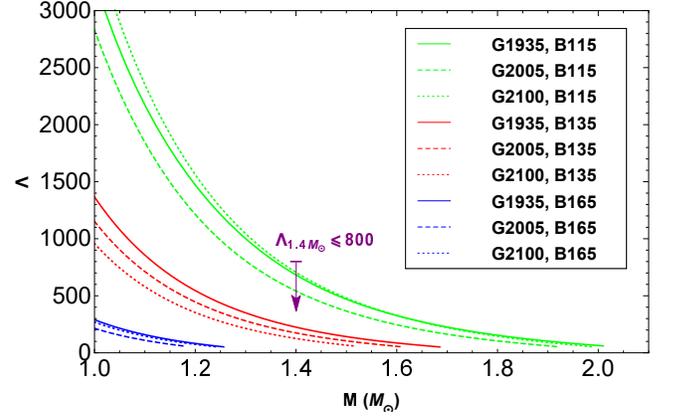}
\caption{Tidal deformability of nonstrange quark stars for the
nine representative EOSs. The constraint of $\Lambda(1.4
M_{\odot}) \leq 800 $ from Ref.~\cite{PhysRevLett.119.161101} is
also plot.} \label{Fig:LambdaM}
\end{figure}

\begin{figure}
\includegraphics[width=0.47\textwidth]{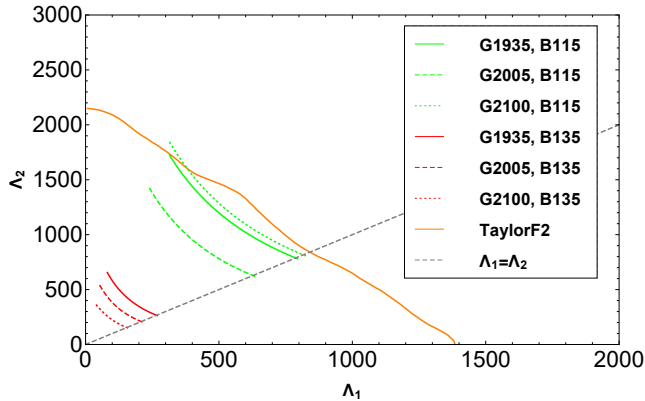}
\caption{$\Lambda_1-\Lambda_2$ relations for the nine
representative EOSs. The observational constraint from GW170817
based on the TaylorF2 waveform model is also plot for comparision
~\cite{PhysRevX.9.011001}.} \label{Fig:Lambda1Lambda2}
\end{figure}

Our numerical results on the tidal deformability of nonstrange
quark stars for the nine representative EOSs are shown in
Fig.~\ref{Fig:LambdaM}. We see that the tidal deformability
decreases monotonously as the mass increases. EOSs with $B^{\frac{1}{4}}=115, 135$ MeV satisfy the constraint of $\Lambda(1.4
M_\odot) \leq 800$ derived from the observations of GW170817 for
the low-spin priors~\cite{PhysRevLett.119.161101}. Furthermore,
based on the waveform model of TaylorF2, the chirp mass
$\mathcal{M} = (M_1 M_2)^{3/5}(M_1+M_2)^{-1/5}$ is restricted to
be $1.186 \pm 0.0001 M_{\odot}$, providing further constraint on
companion mass of $M_1$ and $M_2$ for
GW170817~\cite{PhysRevX.9.011001}. Thus the $\Lambda_1-\Lambda_2$
relation of the BNS can be obtained for our nine representative
EOSs, which is shown in Fig.~\ref{Fig:Lambda1Lambda2}. Here, the
constraint from GW170817 based on the TaylorF2 waveform model
~\cite{PhysRevX.9.011001} is also plot for a direct comparison. We
see that every EOS with $M_{\rm max}\geq1.36 M_{\odot}$ (i.e., the EOSs with $B^{\frac{1}{4}}=115, 135$ MeV) fulfills
the constraint except for the EOS with $G_1=2.1$ GeV$^{-2}$ and
$B^{\frac{1}{4}}=115$ MeV. In addition to that, a larger
$B^{\frac{1}{4}}$ makes the curve closer to the lower left corner of the figure.

For the sake of completeness, the properties of nonstrange quark
stars for the nine representative EOSs are summarized in
Table.~\ref{ninerepresentative}. The parameters listed include the
mass, radius, surface and central energy densities of the heaviest
star, and the tidal deformability and radius of a 1.4 $M_{\odot}$
/ 1.6 $M_{\odot}$ star. From Fig.~\ref{Fig:MR},
Fig.~\ref{Fig:Lambda1Lambda2}, and
Table.~\ref{ninerepresentative}, we can see that the parameter
sets with $B^{\frac{1}{4}}=115$ MeV is good at supporting massive
stars, while the EOSs with $G_1=2.005$ GeV$^{-2}$ correspond to
the original 2-flavor NJL model. Although the EOSs of the normal
NJL model can satisfy the constraints of the tidal deformability
from GW170817, they are not consistent with the mass constraint
from PSR J0740+6620 and the radius constraint from PSR J0030+0451.
In particular, for $B^{\frac{1}{4}}=115$ MeV, both increasing and
decreasing the value of $G_1$ can increase the maximum mass, but a
too large or too small value of $G_1$ makes the EOS conflict with
the tidal deformability measurement of GW170817. Among the nine
representative EOSs, only the one with $B^{\frac{1}{4}}=115$ MeV
and $G_1=1.935$ GeV$^{-2}$ is in agreement with all current
astronomical measurements considered in this study. Based on these
analyses, we argue that nonstrange quark stars might exist in the
universe.

%% It demonstrates that on the one hand, the nonstrange quark star
%% might exist in the universe; on the other hand, the modification
%% of the four-fermion coupling constant $G$ in the NJL model
%% satisfies not only the physical requirements in essence,
%% but also the constraints from astronomical measurements.
%

\begin{widetext}
\begin{center}
\begin{table}
\caption{Properties of nonstrange quark stars for the nine
representative EOSs.}\label{ninerepresentative}
\begin{tabular}{p{1.3cm} p{1.5cm} p{1.0cm} p{1.0cm} p{1.8cm}p{1.8cm}p{1.2cm}p{1.1cm}p{1.1cm}p{1.1cm}}
\hline\hline
$\quad B^{\frac{1}{4}}$&$\quad\,\, G_1$&$M_{\rm max}$&$\,\,R_m$&$\quad\,\,\,\epsilon_0$&$\quad\,\,\,\epsilon_c$&$\Lambda(1.4)$&$R(1.4)$&$\Lambda(1.6)$&$R(1.6)$\\
$\,\,[{\rm MeV}]$&$[{\rm GeV}^{-2}]$&$[M_{\odot}]$&$\,$[km]&$[{\rm MeV/fm^3}]$&$[{\rm MeV/fm^3}]$&$\quad-$&$\,\,$[km]&$\quad-$&$\,\,$[km]\\
\hline
\multirow{3}{*}{$\,\,\,\,$115}&$\,\,\,\, 1.935$&2.10&11.69&$\quad$178&$\quad$1118&$\,\,\,\,$679&11.70&$\,\,\,$324&12.01\\
                              &$\,\,\,\, 2.005$&1.97&11.14&$\quad$188&$\quad$1227&$\,\,\,\,$539&11.35&$\,\,\,$246&11.60\\
                              &$\,\,\,\, 2.100$&2.03&11.62&$\quad$166&$\quad$1113&$\,\,\,\,$702&11.83&$\,\,\,$325&12.10\\
                    \hline
\multirow{3}{*}{$\,\,\,\,$135}&$\,\,\,\, 1.935$&1.74&$\,\,$9.63&$\quad$277&$\quad$1612&$\,\,\,\,$222&$\,\,$9.96&$\,\,\,\,\,$83&10.08\\
                              &$\,\,\,\, 2.005$&1.66&$\,\,$9.25&$\quad$291&$\quad$1756&$\,\,\,\,$176&$\,\,$9.65&$\,\,\,\,\,$59&$\,\,$9.64\\
                              &$\,\,\,\, 2.100$&1.57&$\,\,$8.85&$\quad$305&$\quad$1909&$\,\,\,\,$124&$\,\,$9.31&$\,\,\,\,\,\,-$&$\,\,\,\,\,\,-$\\
                                                  \hline
\multirow{3}{*}{$\,\,\,\,$165}&$\,\,\,\, 1.935$&1.30&$\,\,$7.17&$\quad$513&$\quad$2848&$\,\,\,\,\,\,-$&$\,\,\,\,\,\,-$&$\,\,\,\,\,\,-$&$\,\,\,\,\,\,-$\\
                              &$\,\,\,\, 2.005$&1.23&$\,\,$6.84&$\quad$555&$\quad$3066&$\,\,\,\,\,\,-$&$\,\,\,\,\,\,-$&$\,\,\,\,\,\,-$&$\,\,\,\,\,\,-$\\
                              &$\,\,\,\, 2.100$&1.29&$\,\,$7.09&$\quad$521&$\quad$2860&$\,\,\,\,\,\,-$&$\,\,\,\,\,\,-$&$\,\,\,\,\,\,-$&$\,\,\,\,\,\,-$\\
\hline\hline
\end{tabular}
\end{table}
\end{center}
\end{widetext}

\section{Summary and discussion}
\label{three}

The EOS of a modified 2-flavor NJL model with PTR is introduced to
investigate the properties of nonstrange quark stars. Since the
coupling constant $G$ in the normal NJL model is an indication of
the effective gluon propagator, it should not be ``constant''
according to the simulation of lattice QCD. The corresponding DS
equation should be coupled with that of quarks by QCD in essence.
Inspired by the OPE method, the coupling constant $G$ is modified
as $G = G_1 + G_2 \langle \bar{\psi} \psi \rangle $ in this study,
similar to the treatment in
Refs.~\cite{PhysRevD.85.034031,Cui2013,Cui2014,PhysRevD.93.036006,PhysRevD.94.096003,doi:10.1142/S0217732317501073,Fan_2019,PhysRevD.97.103013}.
The electric charge neutrality and beta
equilibrium are considered in our calculations. Nine
representative EOSs are obtained for nonstrange quark stars,
corresponding to different combinations of $G_1$ and $B$ (the bag
constant) parameters. The sound velocity is calculated to
illustrate the stiffness and rationality of the EOSs. It is found
that the sound velocity is ubiquitously smaller than the speed of
light, not exceeding the conformal limit for dense matter. It is
also found that a larger $G_1$ usually leads to a stiffer EOS.
Interestingly, for the bag constant, a smaller $B$ parameter
yields a softer EOS at low densities; but at high densities, the
case is opposite.

The $M$-$R$, $M$-$\epsilon_{\rm c}$ relations and the tidal
deformability of quark stars are calculated based on our new EOSs.
The results are directly compared with recent astronomical
observations, including the mass measurements of PSR J0740+6620
($2.14^{+0.10}_{-0.09}
M_{\odot}$)~\cite{cromartie2020relativistic} and PSR J0348+0432
($2.01\pm0.04 M_{\odot}$)~\cite{Antoniadis1233232}, the radius
measurements by NICER ($R_{1.44 M_{\odot}}>10.7$
km~\cite{bogdanov2019constraining}, $R_{1.4
M_{\odot}}=11.0^{+0.9}_{-0.6}$ km~\cite{capano2020stringent}, $M =
1.34^{+0.15}_{-0.16}$ $M_{\odot}$ and $R=12.71^{+1.14}_{-1.19}$ km
for PSR J0030+0451~\cite{riley2019nicer}), and the tidal
deformability constraints from
GW170817~\cite{PhysRevLett.119.161101,PhysRevX.9.011001}. It is
found that when the two key parameters are taken as $G_1=1.935$
GeV$^{-2}$ and $B^{\frac{1}{4}}=115$ MeV, the corresponding EOS
can satisfy all the observational constraints listed above,
strongly supporting the possible existence of nonstrange quark
stars. In this case, the nonstrange quark star can have a maximum
mass of $2.1 M_{\odot}$, with a radius of 11.69 km.
%% It demonstrates that the modification of the
%% coupling constant $G$ in the normal NJL model
%% is helpful, because it is not only consistent with
%% the physical requirement in essence, but also in
%% agreement with astronomical observations.

In the future, more and more astronomical
measurements will be available, which can help better constraining
the EOS of QCD. Note that only the relatively simple case of
four-fermion interactions is considered here. But the method could
also be extended to include more interactions, such as the 't
Hooft and eight-quark interactions. To do that, we need to
consider how to deal with their coupling consistently. For the
normal SU(3) case, more free parameters (the current masses of u,
d, s quarks, the coupling constants of four-, six-, eight-quark
interactions, and the ultraviolet cut-off) would be introduced to
characterize the six- and eight-quark interactions in the NJL
model. They should be determined through various laboratory
experiments and astronomical observations, and should be
considered in future theoretical studies.

\acknowledgments

This work is supported in part by the National Natural Science
Foundation of  China (under Grants No. 12005192, No. 12103047, No. 12233002, No.
11873030, No. 12041306, No. 12147103, and No. U1938201), the
Project funded by China Postdoctoral Science Foundation
(2020M672255, 2020TQ0287), National SKA Program of China No.
2020SKA0120300, the National Key R$\&$D Program of China
(2021YFA0718500), the science research grants from the China
Manned Space Project with NO. CMS-CSST-2021-B11, the Natural
Science Foundation of Henan Province of China (212300410290), the
start-up funding from Zhengzhou University, and the High-level
Talent Fund of Henan University of Technology (Grant No.
31401242).

\bibliography{reference}
\end{document}